\documentclass[letterpaper,twocolumn,prl,aps,showpacs,floatfix, 
superscriptaddress]{revtex4-2}
\usepackage[latin1]{inputenc}
\usepackage{bm}
\usepackage[usenames]{color}
\usepackage{multirow}
\usepackage{amssymb}
\usepackage{amsbsy}
\usepackage{amsmath}
\usepackage{stmaryrd}
\usepackage{graphicx}
\usepackage{epsfig}
\usepackage{placeins}
\usepackage{bbold}
\usepackage{bbm}
\usepackage{braket}
\usepackage[colorlinks,linkcolor=blue,citecolor=blue,urlcolor=blue]{hyperref}

\usepackage{scalerel}
\usepackage{tikz}
\usetikzlibrary{calc}
\usetikzlibrary{patterns}
\usetikzlibrary{svg.path}
\definecolor{orcidlogocol}{HTML}{A6CE39}
\tikzset{
  orcidlogo/.pic={
    \fill[orcidlogocol] 
svg{M256,128c0,70.7-57.3,128-128,128C57.3,256,0,198.7,0,128C0,57.3,57.3,0,128,
0C198.7,0,256,57.3,256,128z};
    \fill[white] svg{M86.3,186.2H70.9V79.1h15.4v48.4V186.2z}
                 
svg{M108.9,79.1h41.6c39.6,0,57,28.3,57,53.6c0,27.5-21.5,53.6-56.8,
53.6h-41.8V79.1z 
M124.3,172.4h24.5c34.9,0,42.9-26.5,
42.9-39.7c0-21.5-13.7-39.7-43.7-39.7h-23.7V172.4z}
                 
svg{M88.7,56.8c0,5.5-4.5,10.1-10.1,10.1c-5.6,0-10.1-4.6-10.1-10.1c0-5.6,4.5-10.1
,10.1-10.1C84.2,46.7,88.7,51.3,88.7,56.8z};
  }
}

\newcommand\orcid[1]{\!%
  \href{https://orcid.org/#1}{%
    \mbox{%
      \scaleto{%
        \begin{tikzpicture}[yscale=-1,transform shape]
          \pic{orcidlogo};
        \end{tikzpicture}
      }{8pt}%
    }%
  }%
}

\makeatletter

\begin{document}
\title{Simulating the dynamics of large many-body quantum systems with 
Schr\"odinger-Feynman techniques}

\author{Jonas Richter~\orcid{0000-0003-2184-5275}}
\affiliation{Department of Physics, Stanford University, Stanford, CA 94305, 
USA}
\affiliation{Institut f\"ur Theoretische Physik, Leibniz 
Universit\"at Hannover, 30167 Hannover, Germany}

\date{\today}

\begin{abstract}

The development of powerful numerical techniques has drastically improved 
our understanding of quantum matter out of equilibrium. Inspired by recent 
progress in the area of noisy intermediate-scale quantum devices, this paper  
highlights hybrid Schr\"odinger-Feynman techniques as an innovative approach to 
efficiently simulate certain aspects of many-body quantum dynamics on classical 
computers. To this end, we 
explore the nonequilibrium dynamics of two large subsystems, which interact 
sporadically in time, but otherwise evolve independently from each other. 
We consider subsystems with tunable disorder strength, 
relevant in the context of many-body localization, where one subsystem can 
act as a bath for the other. Importantly, studying the full interacting 
system, we observe that signatures of thermalization are enhanced compared to 
the reference case of having two independent subsystems. Notably, with the 
here proposed Schr\"odinger-Feynman method, we are able to simulate the 
pure-state survival probability in systems  significantly larger than accessible by    
standard sparse-matrix techniques.

\end{abstract}

\maketitle


{\it Introduction.--}
Studying the dynamics of many-body quantum systems out of equilibrium is highly 
challenging. This is not least due to the exponentially growing
Hilbert space with system size and the build-up of entanglement during the 
unitary time evolution. While analytical solutions are typically rare, a 
variety of 
sophisticated numerical 
methods have been developed to 
counteract these challenges. These include, for instance, 
Krylov subspace techniques \cite{Nauts_1983, Long_2003}, dynamical mean field theory \cite{Aoki_2014}, quantum Monte-Carlo \cite{Goth_2012}, classical phase-space representations \cite{Wurtz_2018},
neural network approaches \cite{Carleo_2017, Schmitt_2018, Schmitt_2020}, numerical-linked cluster expansions \cite{White2017, Mallayya_2018, Guardado_Sanchez_2018, Richter_2019} and, last but not least, tensor-network 
methods such as the time-dependent density-matrix renormalization group \cite{Vidal_2004, White_2004, Haegeman_2011, Paeckel_2019}.  
Notwithstanding the immense progress that has been achieved, various questions 
still remain difficult to address, e.g., distinguishing thermalization and 
localization in disordered systems \cite{Weiner_2019, _untajs_2020, _untajs_2020_2, Kiefer_Emmanouilidis_2021, Panda_2020, Sierant_2020, Sels_2021, Abanin_2021}, obtaining quantitative values of diffusion 
constants \cite{White_2018,Ye_2020,Rakovszky_2022, Klein_Kvorning_2022}, or simulating the dynamics of long-range or higher-dimensional 
systems \cite{Zaletel_2015, Kloss_2019, Schuckert_2020, Hubig_2019, Czarnik_2019, Kloss_2020,Richter_2020, Kshetrimayum_2020, Richter_2023}. 

Progress in condensed-matter and quantum many-body physics has been 
influenced substantially by concepts from quantum information. Most notably, 
the notion of entanglement has revolutionized our understanding of phases of 
matter and the complexity of quantum states \cite{Osborne_2002, Schollw_ck_2005, Eisert_2010}. With the recent advent of noisy 
intermediate-scale quantum 
(NISQ) devices \cite{Preskill_2018}, we can witness another example of such mutual influence between 
different subfields. On one hand, the natural language of quantum computers is 
given in terms of quantum circuits, consisting of layers of local gates \cite{Nielsen_2012}. 
Applied to questions in quantum dynamics, this framework of quantum circuits  
has lead to new insights into thermalization, quantum chaos, information 
spreading, and the discovery of exotic out-of-equilibrium phases of 
matter \cite{Chan_2018, Bertini_2019, Claeys_2021, Potter_2022, Lunt_2022, Li_2018, Skinner_2019, Zhou_2020, von_Keyserlingk_2018, Nahum_2018, Khemani_2018, Fisher_2023, Mi_2021}. On the other 
hand, with NISQ devices now featuring a nontrivial number of qubits \cite{Arute_2019, Mi_2021, Kim_2023,Keenan_2023}, 
tailored numerical methods are being developed in order to benchmark the 
quantum simulations \cite{Pednault2017, Huang_2021, Gray_2021,Pan_2022, Pan_2022_2,  Ayral_2023, Tindall2023, Liao2023, Begu_i__2024, Rudolph2023, Anand2023, Kechedzhi_2024}. 
 
One such NISQ-inspired simulation method is given by the Schr\"odinger-Feynman 
approach \cite{Aaronson2016, Chen_2018, Markov2018, Burgholzer_2021}, which combines Schr\"odinger-style evolution of the wave function 
with Feynman-style path summation in a memory-efficient way, allowing the 
simulation of systems out of reach for established sparse-matrix techniques.
In a nutshell, the system of interest is split into smaller patches 
(subsystems) which are simulated independently from each other, thereby reducing 
the overall memory requirements of the simulation. However, with each gate that 
connects two different subsystems, an exponentially growing number 
of trajectories needs to be simulated to recover the dynamics of the full 
system. Thus, Schr\"odinger-Feynman simulations entail favorable memory 
performance 
at the cost of 
an increased overall run time, where the latter can be mitigated by using 
large-scale parallelization of different trajectories.   

While Schr\"odinger-Feynman simulations have been employed to benchmark 
Google's famous ``quantum supremacy'' experiment \cite{Arute_2019}, they have not   
attracted interest yet for questions in many-body dynamics. Here, we 
demonstrate 
their usefulness by  
studying the nonequilibrium dynamics of two large subsystems, which interact 
sporadically in time, but otherwise evolve independently from each other. 
While this setup is interesting in general, one particular motivation for 
us stems from the question of many-body localization (MBL) in disordered 
systems coupled to a bath \cite{Gopalakrishnan_2015, Nandkishore_2015, De_Roeck_2017,  Goihl_2019, Sels_2022, Morningstar_2022, Peacock_2023}. 
To this end, we focus on the pure-state survival probability which we show to 
be 
an especially amenable quantity in Schr\"odinger-Feynman simulations. 
Using only fairly standard computational resources, we are able to 
study the dynamics of rather large systems with up to  
$48$ spin-$1/2$ degrees of freedom. In particular, for these large interacting 
system, we observe that signatures of thermalization are 
enhanced compared to the reference case of having two smaller independent 
subsystems. 

{\it Schr\"odinger-Feynman simulations.--}
On one hand, in the Schr\"odinger (or state-vector) approach \cite{De_Raedt_2007, De_Raedt_2019}, the pure state of 
the system is evolved in time, i.e., $\ket{\psi(t)} = e^{-iHt}\ket{\psi}$ in 
case of Hamiltonian dynamics or, more generally, $\ket{\psi(t)} = 
V(t)\ket{\psi}$ with some unitary $V(t)$, e.g., resulting from a quantum 
circuit. This requires to keep the full state $\ket{\psi(t)}$, i.e., 
exponentially many complex coefficients, in memory. On the other hand, in the 
Feynman approach \cite{Boixo2017}, the final state (or one of its amplitudes) is obtained by 
summing up the contributions of different histories, e.g., $|\langle 
0|V|0\rangle|^2 = |\sum_{\bf x} \langle 0|V_m|x_{m-1} \rangle \cdots 
\langle x_2|V_2|x_1\rangle \langle x_1| V_1|0\rangle|^2$, where we have written 
$V = V_m \cdots V_2 V_1$ as a product of elementary (e.g.,\ two-qubit) gates 
and the sum is over all combinations of intermediate computational 
basis states. In 
contrast to the state-vector approach, the Feynman technique is memory 
efficient as we only need to track the transitions 
between different basis 
states under the action of the two-qubit gates $V_i$. 
However, the simulation time (i.e., the number of possible histories) grows 
exponentially with depth $m$. The goal of the
here employed Schr\"odinger-Feynman method is to combine 
these two paradigms 
to achieve both favorable time and memory requirements \cite{Aaronson2016, Chen_2018, Markov2018, Burgholzer_2021}.
\begin{figure}[tb]
 \centering
 \includegraphics[width=\columnwidth]{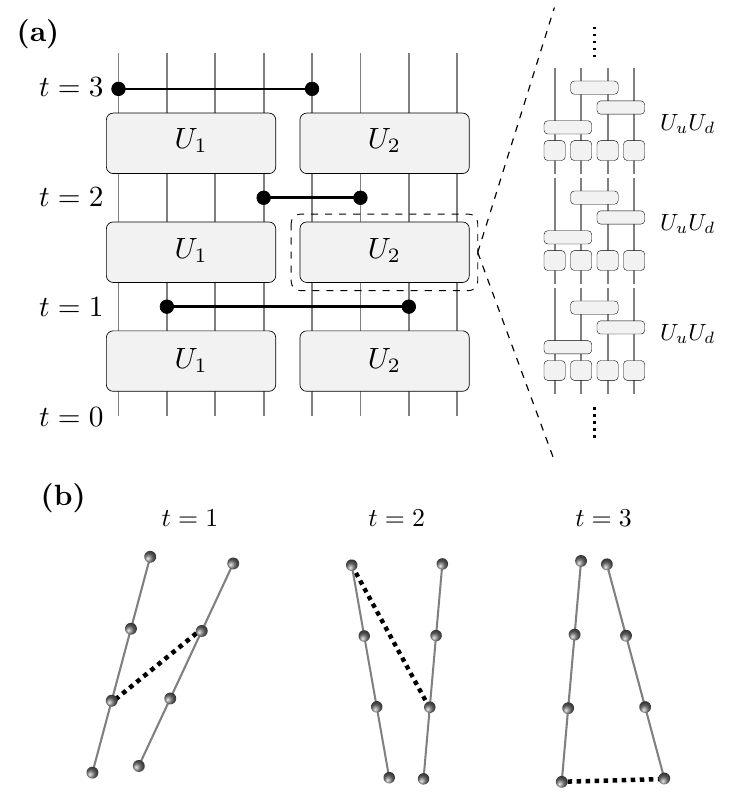}
 \caption{Sketch of the setup studied in this paper. \textbf{(a)} Two 
subsystems evolve independently from each other with 
respect to unitaries $U_{1(2)}$, obtained from 
an elementary Floquet unitary with tunable disorder strength \cite{Morningstar_2022}, e.g., $U_2 = (U_uU_d)^{N_p}$, applied 
$N_p$ times (see text for more details). Then 
a two-qubit 
gate, for instance a CZ gate, is applied to connect the subsystems. This 
completes one time step. The 
two positions used in different time steps for 
the connecting two-qubit gate 
vary randomly over the whole size of the subsystems. 
\textbf{(b)} We consider 
one-dimensional subsystems for simplicity. 
The dashed lines indicate examplary 
positions used for the connecting gate 
in different time steps. Our numerical results 
are obtained by averaging over 
random realizations of the 
initial product state, the unitaries $U$ 
(independent for each subsystem), and 
the position of the connecting two-qubit gate.}
 \label{Fig1}
\end{figure}

The central idea is to perform ``circuit-cutting'' (cf.\ \cite{Peng_2020, Barratt_2021, Bravyi_2016_2, Mitarai_2021, Gentinetta_2024}), i.e., dividing the whole system into smaller patches whose time 
evolutions are simulated independently of each other by a full 
Schr\"odinger approach.  The individual patches should be chosen 
as large as possible in order to efficiently utilize the available memory. 
Anticipating our numerical example below, 
let us consider for 
concreteness a system with a total of $L$ spin-$1/2$ degrees 
of freedom (i.e., 
qubits). The memory requirements to store the full 
state $\ket{\psi(t)}$ would scale as $\sim 2^L$. On the other 
hand, splitting 
the system in two halves and simulating the subsystems individually, memory 
would only scale as $\sim 2\times 2^{L/2} = 2^{L/2+1}$. 

The splitting of the system into patches naturally comes at a price. 
In particular, 
for each gate that connects 
different patches, the number of independent ``trajectories'' grows exponentially. (This is reminiscent of Clifford+T gate simulators, where the computational costs grow exponentially with the number of non-Clifford gates \cite{Bravyi_2016}.)  
Specifically, consider a system split in two halves, which interact 
via a two-qubit gate where one qubit 
is chosen from the first subsystem and the other qubit is chosen from the 
second subsystem.
In general, such a two-qubit gate is a $4\times 4$ matrix which can be
decomposed as \cite{Markov2018}, 
\begin{align}
 \begin{pmatrix} 
 A_{00} & A_{01} \\
  A_{10} & A_{11}
  \end{pmatrix} &= P_0 \otimes A_{00} + 
  P_1 \otimes A_{11} \nonumber  \\ &+ \ket{0}\bra{1} 
\otimes A_{01} + \ket{1}\bra{0} 
\otimes A_{10}\ , \label{Eq::GeneralGate}
\end{align}
where $P_0 = \ket{0}\bra{0}$ and $P_1=\ket{1}\bra{1}$ are 
projections on the 
local basis states of the first qubit, and the $A$ are 
$2\times 2$ matrices (i.e., general one-qubit gates) acting on the qubit in the 
second subsystem. Note that we could have also switched 
the roles of subsystem one and two in this decomposition. For a given trajectory, a single term from the right hand side of Eq.\ \eqref{Eq::GeneralGate} is applied, and the full action of the gate recovered by suitably combining all possible trajectories (see below). Thus, in the most 
general case, the number of trajectories to be simulated is $4^t$ where $t$ is the number 
of connecting two-qubit gates (we choose $t$ as the unit of time in our 
numerical examples, see also Fig.\ \ref{Fig1}).   

In this paper, we will consider two specific examples of two-qubit gates 
connecting different patches, which are relevant to NISQ applications. 
The first example is a controlled-$Z$ gate,
\begin{equation}\label{Eq::CZGate}
 \text{CZ} = \text{diag}(1,1,1,-1) = P_0 \otimes \mathbb{1} + P_1 \otimes Z\ , 
\end{equation}
where $Z$ is a Pauli matrix. The decomposition on the right hand side of Eq.\ 
\eqref{Eq::CZGate} indicates that the simulation of subsystems 
interacting via \text{CZ} gates is less complex than the general case in Eq.\ 
\eqref{Eq::GeneralGate}. In particular, the number of trajectories to be 
simulated is only given by $2^{\# \text{CZ}}$. In practice, we will 
use parallelization to simulate trajectories on multiple processors. A 
processor is given one of $2^{\# \text{CZ}}$ bitstrings, which encodes a 
particular trajectory. 

As a more challenging example, we will also consider subsystems connected by 
\text{iSWAP} gates,
\begin{align}
 \text{iSWAP} &= P_0 \otimes P_0 + P_1 \otimes P_1 \nonumber \\
 &+ i\ket{0}\bra{1} \otimes \ket{1}\bra{0}  + i\ket{1}\bra{0} \otimes 
\ket{0}\bra{1}\ , \label{Eq::iSwapGate}
\end{align}
for which the number of trajectories is $4^{\# \text{iSWAP}}$.

The total state is obtained by 
combining the contributions 
in a Feynman-like fashion,
\begin{equation}\label{Eq::TotalState}
 \ket{\psi} = \sum_{k,l} \psi_{kl} \ket{k}  \ket{l} =  
\sum_{k,l} \sum_{{\bf j}} \psi_k^{({\bf j})} \psi_l^{({\bf j})} 
\ket{k_{\bf j}} \ket{l_{\bf j}}\ , 
\end{equation}
where the \textbf{j}-sum runs over all independent trajectories, and the 
$k,l$-sums run over the $2^{L/2}$ basis states in the two subsystems 
respectively. As the states of the subsystems are evolved in 
time \textit{\`a 
la} Schr\"odinger, the coefficients $\psi_k^{({\bf j})}$ and $\psi_l^{({\bf 
j})}$ are generally kept 
in memory, whereas the $2^L$ coefficients of the full state 
$\ket{\psi}$ typically entail prohibitive memory requirements. Therefore, it might be necessary to 
additionally store the $\psi_k^{({\bf j})}$ and $\psi_l^{({\bf 
j})}$ to hard drive, and construct the coefficients of 
$\ket{\psi}$ required to evaluate observables after the simulation.  

Generalizing Eq.\ \eqref{Eq::TotalState} to more than two patches is 
straightforward. Moreover, if a nonperfect accuracy is acceptable, 
computational resources can be saved by simulating only a smaller (randomly 
chosen) fraction of trajectories 
in Eq.\ \eqref{Eq::TotalState}. For more details on Schr\"odinger-Feynman 
simulations and state-of-the-art 
implementations, see \cite{Markov2018, Chen_2018}.

While impressive quantum-circuit simulations have been performed \cite{Arute_2019, Markov2018, Chen_2018}, it might be fair to say that the Schr\"odinger-Feynman method is less known in the qunatum many-body physics community. Bridging this gap is a goal of the present work. 

{\it Model and Observables.--}
A natural application of the Schr\"odinger-Feynman technique is a 
situation, where two relatively large (sub-)systems interact weakly with each 
other. 
For instance, one could think of a system that interacts with a bath or 
reservoir (although, here, the bath will not be infinite or even 
bigger than the system), but the system-bath interaction only occurs 
sporactically in time. While this 
setup might seem somewhat artificial, a similar model was recently 
considered in Ref.\ \cite{Peacock_2023}, where a strongly and a weakly disordered subsystem 
were brought into contact to study the effect of a thermal inclusion on 
the putative many-body localization transistion. Our model studied below is 
partially inspired by Ref.\ \cite{Peacock_2023}, although we stress that we here do not aim 
to quantitatively address the stability of MBL. 

Our setup is sketched in Fig.\ \ref{Fig1}. Specifically, we consider a system split into two subsystems. Each 
subsystem is chosen as a linear chain of length $L_1$, $L_2$ such that the 
total size of the 
system is $L = L_1 + L_2$ (we actually consider $L_1 = L_2 = L/2$). The 
chain geometry is chosen for 
simplicity, but does not represent a conceptual limitation of the 
Schr\"odinger-Feynman technique. Each subsystem evolves individually in time 
with respect to a unitary that can be tuned between a weak and a strong 
disorder regime. At discrete points in time, both subsystems interact with each 
other via a \text{CZ} or \text{iSWAP} gate, where the two qubits 
in the two subsystems are chosen randomly at each point in time.  

For the unitary evolution, we study a variant of the Floquet random-circuit 
model introduced in Ref.\ \cite{Morningstar_2022} in the context of many-body localization.
One Floquet period is given by a unitary
$U = U_u U_d$, where $U_d$ is build from one-qubit gates, 
$U_d = d_1 \otimes d_2 \otimes \cdots \times d_L$. Each $d_\ell$ is a diagonal 
matrix, obtained by drawing a $2\times 2$ random matrix (different for 
each subsystem and site $\ell$) from the circular unitary ensemble and 
diagonalizing it. The 
computational $Z$ basis thus corresponds to the eigenbasis of the 
$d_\ell$. Further, $U_u$ consists 
of 
nearest-neighbor two-qubit gates $u_\ell = \exp(iM_\ell/\alpha)$, which are applied in a randomly chosen sequence, 
$U_u = \Pi_\ell u_{\pi(\ell)}$, where $\pi$ denotes a permutation and $M_\ell \in 
\mathbb{C}^{4\times 4}$ is drawn from the Gaussian unitary ensemble.  
The disorder strength is controlled by the choice of $\alpha$, where a large 
$\alpha$ corresponds to strong disorder (one-qubit gates dominate), and a small 
$\alpha$ corresponds to weak disorder (two-qubit gates dominate).
For more details on the MBL random-circuit model, see Refs.\ \cite{Morningstar_2022, Ha_2023}.

In practice, we will generate 
random unitaries $U_1$ and $U_2$ (different for 
each subsystem), with potentially different disorder strength $\alpha_1$ and 
$\alpha_2$ and differerent orders of the two-qubit gates in $U_u$. The 
subsystems evolve individually for a number of Floquet periods before 
interacting via 
a $\text{CZ}$ or $\text{iSWAP}$ gate. That is, we will define one time step 
below as, e.g., $\ket{\psi(t+1)} = \text{CZ} U_1 U_2 \ket{\psi(t)}$, where 
$U_1 = (U_{u,1}U_{d,1})^{N_p}$. The number of Floquet periods within the 
unitaries $U_1$ and $U_2$ are fixed to $N_p = 10$ to mimic that the 
subsystems are strongly interacting and building up entanglement internally, while only weakly interacting 
with each other.

Our results are averaged over multiple 
random-circuit realizations with independently chosen 
$U_d$ and $U_u$ in each run, as well as randomly chosen positions of the qubits 
involved in the interactions between subsystems [Fig.\ \ref{Fig1}~(b)], i.e., interpreting our setup 
again as a system and bath, the interactions with the bath can occur globally, 
but only sporadically in time. 
Let us also emphasize that the Schr\"odinger-Feynman approach is 
agnostic to the specific type of unitary time evolution. Instead of the 
random-circuit model, we could have equally well considered Hamiltonian 
time evolution of the two subsystems. 

As an obervable, we consider the survival probability of some 
out-of-equilibrium  
initial state $\ket{\psi(0)}$ \cite{Torres_Herrera_2015},
\begin{equation}\label{Eq::LT}
 {\cal L}(t) = \left|\braket{\psi(0)|\psi(t)}\right|^2\ .
\end{equation}
The initial state is chosen as a product state from the $Z$ basis, i.e., 
$\ket{\psi(0)} = \ket{k}\ket{l}$ [in the notation of Eq.\ 
\eqref{Eq::TotalState}] with randomly chosen product states $\ket{k}$ and $\ket{l}$. Then, crucially, in order to evaluate ${\cal L}(t)$ in 
Eq.\ \eqref{Eq::LT}, only this single 
amplitude of the full time-evolved state $\ket{\psi(t)}$ is required, i.e., 
${\cal L}(t) = |\sum_\textbf{j} \psi_k^{(\textbf{j})}(t) 
\psi_l^{(\textbf{j})}(t)|^2$ with $k,l$ fixed by the initial state. The subsystem-state coefficients $\psi_k^{(\textbf{j})}(t)$ and $\psi_l^{(\textbf{j})}(t)$, distributed over multiple processors which each handle a particular trajectory \textbf{j}, are 
collected at each time step and ${\cal L}(t)$ is calculated during runtime. 
Thus, saving the states of the subsystems to hard drive is not required, which 
makes ${\cal L}(t)$ particularly suitable for the Schr\"odinger-Feynman 
technique. In our simulations, we average ${\cal L}(t)$, as mentioned above, 
over different random-circuit realizations, as well as random choices of the 
initial product state $\ket{\psi(0)}$.   
\begin{figure}[tb]
 \centering
 \includegraphics[width=\columnwidth]{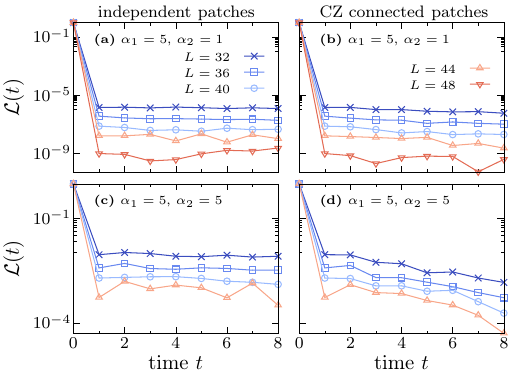}
 \caption{${\cal L}(t)$ in circuits of size $L = L/2 + L/2$. {\bf [(a),(b)]} The first subsystem is strongly disordered with $\alpha_1 = 5$ while the second subsystem is weakly disordered with $\alpha_2 = 1$. {\bf [(c),(d)]} Both subsystems are strongly disordered with $\alpha_{1,2} = 5$. Panels (a) and (c), i.e., {\it left} column, show ${\cal L}(t)$ for the reference case of two independent subsystems (no connecting gates), while panels (b) and (d), i.e., {\it right} column show dynamics in the full interacting system with sporadic CZ gates connecting the two patches. The data for larger $L$ shows stronger fluctuations as it is averaged over fewer random-circuit realizations.}
 \label{Fig_CZ}
\end{figure}

{\it Results.--}
We now present our numerical results. In Fig.\ \ref{Fig_CZ}, the decay of ${\cal L}(t)$ is shown for two subsystems of size $L/2$ connected by sporadic CZ gates up to $t \leq 8$ (cf.\ Fig.\ \ref{Fig1} for the definition of one time step). We consider two different cases: (i) the dynamics of the first patch is strongly disordered with $\alpha_1 = 5$, while the second patch is only weakly disordered with $\alpha_2 = 1$ [Fig.\ \ref{Fig_CZ}~(a) and (b)]; (ii) both subsystems are strongly disordered with $\alpha_{1,2} = 5$ [Fig.\ \ref{Fig_CZ}~(c) and (d)]. Moreover, in both cases, we compare ${\cal L}(t)$ in the fully interacting system, cf.\ Fig.\ \ref{Fig_CZ}~(b) and (d), to the reference case of having two independent subsystems, i.e., by removing the connecting CZ gates, cf.\ Fig.\ \ref{Fig_CZ}~(a) and (c).

Starting from ${\cal L}(0) = 1$, we observe that most of the decay of ${\cal L}(t)$ happens during the first time step $t \leq 1$. Moreoever, comparing different system sizes up to $L \leq 48$, we find that especially for the weakly disordered case with $\alpha_2 = 1$, ${\cal L}(t)$ decreases approximately exponentially with $L$ as expected given the exponentially growing Hilbert space.

While most of the dynamics of ${\cal L}(t)$ is thus caused by the internal dynamics within the subsystems, we find that the sporadic interaction between the subsystems leads to a further slow decay of ${\cal L}(t)$ at $t > 1$. This can be seen especially for $\alpha_1 = \alpha_2 = 5$, where ${\cal L}(t)$ is essentially time-independent in the case of two disconnected patches [Fig.\ \ref{Fig_CZ}~(c)], while a monotonous decay of ${\cal L}(t)$ persists if the patches are connected to form a larger system [Fig.\ \ref{Fig_CZ}~(d)]. The coupling of the two subsystems mediated by the CZ gates thus allows $\ket{\psi(t)}$ to explore a larger Hilbert space even though the disorder strength is the same in both patches. This is consistent with the fact that signatures of thermalization (or potential localization) in disordered systems are subject to pronounced finite-size effects and require the simulation of large systems \cite{Weiner_2019, _untajs_2020, _untajs_2020_2, Kiefer_Emmanouilidis_2021, Panda_2020, Sierant_2020, Sels_2021, Abanin_2021}.
\begin{figure}[b]
 \centering
 \includegraphics[width=\columnwidth]{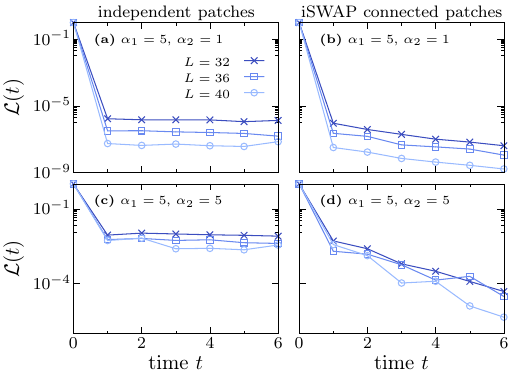}
 \caption{Analogous data as in Fig.\ \ref{Fig_CZ}, but now the two subsystems are connected by sporadic iSWAP gates.}
 \label{Fig_ISWAP}
\end{figure}

Eventually, Fig.\ \ref{Fig_ISWAP} presents analogous data, but now for subsystems connected by iSWAP gates. Due to their higher complexity, cf.\ Eq.\ \eqref{Eq::iSwapGate}, we only simulate dynamics in systems with $L \leq 40$ and up to a depth of $t \leq 6$, which corresponds to $4^6 = 2^{12}$ distinct trajectories, 
Comparing Fig.\ \ref{Fig_CZ} and Fig.\ \ref{Fig_ISWAP}, it appears 
that using iSWAP gates leads to stronger ``interactions'' between 
the two subsystems as the monotonous decay of ${\cal L}(t)$ for $t \geq 1$ in Figs.\ \ref{Fig_ISWAP}~(b) and (d) is now even more pronunced. 
This can be understood from a physical point of 
view. In particular, \text{CZ} and \text{iSWAP} 
gates qualitatively differ from each in the sense that iSWAP gates can change the expectation values $\langle Z \rangle_1$ and $\langle Z \rangle_2$ of the total magnetization in the two subsystems, whereas $\langle Z\rangle_{1,2}$ is conserved under the application of a CZ gate. Note that in our numerical example, the unitaries $U_{1,2}$ in the Floquet circuit do not conserve $\langle Z \rangle_{1,2}$. However, one could imagine a similar random circuit with individual $U(1)$ symmetries in the two subsystems, cf.\ Ref.\ \cite{Jonay2022}, or generate the dynamics instead by $e^{-iHt}$ with $H$ the standard MBL XXZ chain which is also $U(1)$-symmetric.
For such cases, we expect that the thermalization-benefiting effect of the iSWAP gates would become even more pronounced (in comparison to the subsystem-particle-number-conserving CZ gates). 

Figures \ref{Fig_CZ} and \ref{Fig_ISWAP} exemplify that the Schr\"odinger-Feynman technique allows us to study ${\cal L}(t)$ in systems with $L\leq 48$ with a Hilbert-space dimension of $\sim 10^{14}$, where we used parallelization with up to $256$ processes in an MPI architecture on a readily available central university computer cluster. In principle, from a memory and CPU number point of view, even larger systems and longer times would have been accessible for us. Note, however, that the here performed averaging over disorder realizations and initial states added another layer of computational complexity. Moreover, with comparable computational resources, standard 
sparse-matrix techniques would be restricted to systems of roughly half the size reached here. 

{\it Conclusions.--}
We have demonstrated the applicability of Schr\"odinger-Feynman simulations for 
questions in many-body quantum dynamics. Specifically, we have explored the 
decay of the pure-state survival probability in a 
disordered model (here in the form of a Floquet random circuit), where 
the strength of disorder is tunable such that one subsystems can 
act as a thermal bath for the other, cf. Ref.\ \cite{Peacock_2023}. Studying large subsystems that interact sporadically in time, we have observed that signatures of thermalization become 
enhanced compared to the reference case of having two independent patches. 

The system sizes of $L \leq 48$ reported here by no means represent 
the upper limit that can be managed by Schr\"odinger-Feynman techniques, see e.g., Ref.\ \cite{Chen_2018}. With the 
ever increasing availability of memory and processors, it should be 
possible in the future to push these methods to even larger subsystems with 
more connecting gates.   

While our work is supposed to be a proof-of-principle demonstration, we hope 
that it 
will motivate the usage of 
Schr\"odinger-Feynman techniques in the quantum-dynamics community. Promising 
settings, as hinted at in this paper, are subsystems that are highly entangled internally, yet only interact weakly with each other, e.g., systems embedded in a bath, 
where the weak system-bath coupling can effectively be modelled by interactions that occur 
sparsely in time, which might also include certain impurity problems \cite{Sun_2016}. Moreover, while we here considered one-dimensional 
subsystems, more complicated geometries, for which no other efficient numerical methods may exist, can readily be treated within the same framework. 

We thank Yaodong Li for useful discussions. J.\,R.\ acknowledges funding from the European Union's Horizon Europe research 
and innovation programme, 
Marie Sk\l odowska-Curie grant no.~101060162, and the Packard Foundation
through a Packard Fellowship in Science and Engineering (V.\, Khemani's grant). The simulations were carried out on Leibniz Universit\"at Hannover's computer cluster, funded by the Deutsche Forschungsgemeinschaft (DFG, German Research Foundation) -- no.~INST 187/592-1 FUGG, INST 187/742-1 FUGG.


%

\end{document}